\documentclass[11.5pt]{article}
\usepackage{amsfonts}
\usepackage{mathrsfs}
\usepackage{amsmath,amssymb}
\usepackage{latexsym}
\usepackage{graphicx}
\usepackage{color}

\newcommand{\s}{\sigma}
\newcommand{\p}{\partial}

\newcommand{\be}{\begin{eqnarray}}
\newcommand{\ee}{\end{eqnarray}}
\newcommand{\beq}{\begin{equation}}
\newcommand{\eeq}{\end{equation}}
\newcommand{\bee}{\begin{eqnarray*}}
\newcommand{\eee}{\end{eqnarray*}}

\newtheorem{thm}{Theorem}[section]
\newtheorem{lem}{Lemma}[section]

\begin{document}

\title{Optimal Investment with Stopping in Finite Horizon\thanks{\noindent  The project is supported by
 NNSF of China (No.11271143 and No.11371155), University Special Research
 Fund for Ph.D. Program of China (20124407110001 and 20114407120008), and Research Grants Council of Hong Kong under grants 521610 and 519913.}}

\author{\normalsize Xiongfei Jian$^{a}$, \quad {Xun Li$^{b}$} \quad and \quad Fahuai Yi$^{a}$\thanks{\noindent Corresponding author. fhyi@scnu.edu.cn. Tel: 86-20-85216655-8214. Fax: 86-20-85212131.} \\
\small $^a$School of Mathematical Sciences, South China Normal University, Guangzhou, China\\
\small $^b$Department of Applied
Mathematics, Hong Kong Polytechnic University, Hong Kong }

\date{}

\maketitle

\begin{abstract}
In this paper, we investigate dynamic optimization problems featuring both
stochastic control and optimal stopping in a finite time horizon.
The paper aims to develop new methodologies, which are
significantly different from those of mixed dynamic optimal control
and stopping problems in the
existing literature, to study a manager's decision.
We formulate our model to a free boundary problem of a fully \textit{nonlinear} equation.
Furthermore, by means of a dual transformation for the above problem,
we convert the above problem to a new free boundary problem of a \textit{linear} equation.
Finally, we apply the theoretical results to challenging,
yet practically relevant and important, risk-sensitive problems in
wealth management to obtain the properties of the optimal strategy and the right time to achieve a certain
level over a finite time investment horizon.

\vspace{0.2cm}


\noindent {\bf MSC Classification(2010):} 35R35; 91B28; 93E20.

\end{abstract}

\vspace{1cm}

\section{Introduction}
\setcounter{section}{1} \setcounter{equation}{0}

Optimal stopping problems, a variant of optimization problems
allowing investors freely to stop before or at the maturity in order
to maximize their profits, have been implemented in practice and
given rise to investigation in academic areas such as science,
engineering, economics and, particularly, finance. For instance,
pricing American-style derivatives is a conventional optimal
stopping time problem where the stopping time is adapted to the
information generated over time. The underlying dynamic system is
usually described by stochastic differential equations (SDEs). The
research on optimal stopping, consequently, has mainly focused on
the underlying dynamic system itself.
In the field of financial
investment, however, an
investor frequently runs into investment decisions where investors
stop investing in risky assets so as to maximize their expected
utilities with respect to their wealth over a finite time investment
horizon. These optimal stopping problems depend on underlying
dynamic systems as well as investors' optimization decisions (controls).
This naturally results in a mixed optimal control and stopping problem, and
Ceci-Bassan (2004) is one of the typical representatives along this line of
research. In the general formulation of such models, the control is mixed,
composed of a control and a stopping time. The theory has also been
studied in Bensoussan-Lions (1984), Elliott-Kopp (1999), Yong-Zhou (1999) and Fleming-Soner (2006),
and applied in finance in
Dayanik-Karatzas (2003), Henderson-Hobson (2008), Li-Zhou (2006), Li-Wu
(2008, 2009) and Shiryaev-Xu-Zhou (2008).

In the finance field, finding an optimal stopping time point has been
extensively studied for pricing American-style options, which allow option
holders to exercise the options before or at the maturity. Typical examples
that are applicable include, but are not limited to, those presented in
Chang-Pang-Yong (2009), Dayanik-Karatzas (2003) and R\"uschendorf-Urusov
(2008). In the mathematical finance literature, choosing an optimal stopping
time point is often related to a free boundary problem for a class of
diffusions (see Fleming-Soner (2006) and Peskir-Shiryaev (2006)).
In many applied areas, especially in more extensive investment
problems, however, one often encounters more general controlled diffusion
processes. In real financial markets, the situation is even more complicated
when investors expect to choose as little time as possible to stop portfolio
selection over a given investment horizon so as to maximize their profits
(see Samuelson (1965), Karatzas-Kou (1998), Karatzas-Sudderth (1999),
Karatzas-Wang (2000), Karatzas-Ocone (2002), Ceci-Bassan (2004), Henderson (2007),
Li-Zhou (2006) and Li-Wu (2008, 2009)).

The initial motivation of this paper comes from our recent studies
on choosing an optimal point at which an investor stops investing
and/or sells all his risky assets (see Choi-Koo-Kwak (2004) and Henderson-Hobson (2008)). The objective is to find an
optimization process and stopping time so as to meet certain
investment criteria, such as, the maximum of an expected utility
value before or at the maturity. This is a typical problem in the
area of financial investment. However, there are fundamental
difficulties in handling such optimization problems. Firstly, our
investment problems, which are different from the classical
American-style options, involve optimization process over the entire
time horizon. Secondly, it
involves the portfolio in the drift and volatility terms so that the
problem including multi-dimensional financial assets is more realistic
than those addressed in finance literature (see Capenter (2000)). Therefore, it is difficult to solve
these problems either analytically or numerically using current
methods developed in the framework of studying American-style options.
In our model, the corresponding HJB equation of the problem is formulated into a variational inequality
of a fully nonlinear equation. We make a dual transformation for the
problem to obtain a new free boundary problem with a linear equation. Tackling this new free boundary problem,
we establish the properties of the free boundary and optimal
strategy for the original problem.

The remainder of the paper is organized as follows. In Section 2, the mathematical formulation of the
model is presented, and the corresponding HJB equation is posed.
In Section 3, a dual transformation converts the free boundary problem of a fully \textit{nonlinear} PDE to
a new free boundary problem of a \textit{linear} equation \textit{but} with a complicated constraint \eqref{4.15}.
In Section 4, it is a further idea that we simplify the constraint condition in \eqref{4.15} to obtain a new
free boundary problem with a \textit{simple} condition \eqref{5.5}. Moreover, we show
that the solution of problem \eqref{5.5} must be the solution of problem \eqref{4.15}.
Section 5 devotes to the study for the free boundary of problem \eqref{5.5}. In Section 6, we go back to the original
problem \eqref{3.2} to show that its free boundary is decreasing and differentiable.
Moreover we present its financial meanings.
Section 7 concludes the paper.

\section{Model Formulation}
\setcounter{section}{2} \setcounter{equation}{0}

\subsection{The manager's problem}
The manager operates in a complete, arbitrage-free, continuous-time
financial market consisting of a riskless asset with instantaneous
interest rate $r$ and $n$ risky assets, The risky asset prices $S_i$ are
governed by the stochastic differential equations
\be\label{eq:S}
\frac{dS_{i,t}}{S_{i,t}}=(r+\mu_i)dt+\s'_{i}\; dW_t^j, \quad \mbox{for } i = 1,2,\cdots,n,
\ee
where the interest rate $r$, the excess appreciation rates $\mu_i$, and the volatility vectors
$\sigma_i$ are constants,
$W$ is a standard $n$-dimensional Brownian motion.
In addition, the covariance matrix $\Sigma=\s'\s$ is strongly nondegenerate.

A trading strategy for the manager is an $n$-dimensional process $\pi_t$ whose $i$-th component, where $\pi_{i,t}$
is the holding amount of the $i$-th risky asset in the
portfolio at time $t$. An admissible trading strategy $\pi_t$
must be progressively measurable with respect to $\{\mathcal{F}_t\}$ such that $X_t\ge0$. Note that
$X_t=\pi_{0,t}+\sum\limits_{i=1}^n\pi_{i,t}$, where $\pi_{0,t}$ is the amount invested
in the money. The value of the wealth $X_t$ evolves according to
\be\label{eq:X}
dX_t=(rX_t+\mu'\pi_t)dt+\pi_t'\s dW_t.
\ee
In addition, short-selling is allowed.

The manager controls assets with initial value $x$.
The manager's dynamic problem is to choose an admissible trading
strategy $\pi_t$ and a stopping time $\tau$ to maximize his expected utility of
the exercise wealth:
\be\label{eq:value}
V(x,t)=\max\limits_{\pi,\tau}\mathbb{E}\big[e^{-r(\tau-t)}U(X_\tau+K)\big],
\ee
where  $r> 0$ is the interest and $K$ is a positive constant (e.g., a fixed salary),
\begin{align*}
U(x)=\frac{1}{\gamma}\;x^{\gamma}, \quad 0<\gamma<1,
\end{align*}
is the utility function.

 \subsection{HJB equation}\label{sec:unconstrained}

Applying dynamic programming principle, we get the following Hamilton-Jacobi-Bellman (HJB) equation
\be\label{eq:V}
\left\{\begin{array}{ll}
\min\Big\{-\p_t V-\max\limits_{\pi}\Big[\frac{1}{2}(\pi'\Sigma\pi )\p_{xx}V+\mu'\pi \p_x V\Big]-rx \p_x V+r V,\; V- \frac{1}{\gamma}(x+K)^{\gamma}\Big\}= 0,\\
\hspace{7cm} \quad x>0,\;0<t<T, \\
V(0,t)=\frac{1}{\gamma}K^{\gamma}, \quad\quad 0<t<T,\\ [2mm]
V(x,T)=\frac{1}{\gamma}(x+K)^{\gamma}, \quad\quad  x>0.
\end{array} \right.
\ee

Suppose that $V(x)$ is strictly increasing and strictly concave, i.e., $\p_x V>0,\ \p_{xx}V < 0$.
Note that the gradient of $\pi'\Sigma\pi$ with respect to $\pi$
$$\nabla_\pi(\pi'\Sigma\pi)=2\Sigma\pi,$$
then
\be\label{eq:pi}
\pi^*=-\Sigma^{-1} \mu\;\frac{\p_x V(x,t)}{\p_{xx}V(x,t)}.
\ee
Thus \eqref{eq:V} becomes
\be\label{3.2}
\left\{\begin{array}{ll}
\min\Big\{-\p_t V+\frac {1}{2}a^2\frac{(\p_x V)^2}{\p_{xx}V}-rx \p_x V+r V,\;\;V- \frac{1}{\gamma}(x+K)^{\gamma}\Big\}=0,\;\;\; x>0,\;0<t<T, \\ [2mm]
V(0,t)=\frac{1}{\gamma}K^{\gamma},\;\;\;\;\;\;\;\;0<t<T,\\ [2mm]
V(x,T)=\frac{1}{\gamma}(x+K)^{\gamma},\;\;\;\;\; x>0,
\end{array}\right.
\ee
where $a^2=\mu'\Sigma^{-1} \mu$. Now we find a condition under which the free boundary exists. A simple calculation shows
\bee
&&U(x+K)=\frac 1\gamma (x+K)^{\gamma},\\
&&\p_x U(x+K)=(x+K)^{\gamma-1},\\
&&\p_{xx} U(x+K)=-(1-\gamma)(x+K)^{\gamma-2}.
\eee
It follows that
\bee
&&-\p_t U(x+K)+\frac 12 a^2\frac{(\p_x U(x+K))^2}{\p_{xx}U(x+K)}-rx \p_x U(x+K)+r U(x+K)\\
&&=-\frac {a^2}{2}\frac 1{1-\gamma}(x+K)^{\gamma}-rx(x+K)^{\gamma-1}
   +\frac r\gamma (x+K)^{\gamma}\\
&&\geq 0.
\eee
Eliminating $\frac 1\gamma(x+K)^{\gamma-1}$ yields
\bee
&&-\frac {a^2\gamma}{2(1-\gamma)}(x+K)-r\gamma x
   + r(x+K) \geq 0,
\eee
i.e.,
\be\label{3.3}
\Big(\frac {a^2\gamma}{2(1-\gamma)}-r+r\gamma\Big)x\leq \Big(-\frac {a^2\gamma}{2(1-\gamma)}+r\Big)K.
\ee
If
\be\label{3.4}
\frac {a^2\gamma}{2(1-\gamma)}-r\leq -r\gamma,
\ee
then \eqref{3.3} holds for any $x>0$, the solution to problem \eqref{3.2} is $U(x+K)$ at all.

If
\be\label{3.5}
\frac {a^2\gamma}{2(1-\gamma)}-r\geq 0,
\ee
then \eqref{3.3} is impossible for any $x>0$. Therefore, in this case, the solution to problem \eqref{3.2} satisfies
\be\label{3.6}
\left\{\begin{array}{ll}
-\p_t V+\frac {a^2}{2}\frac{(\p_x V)^2}{\p_{xx}V}-rx \p_x V+r V=0, \quad x>0,\;0<t<T, \\ [2mm]
 V(0,t)=\frac{1}{\gamma}K^{\gamma}, \quad\quad 0<t<T,\\ [2mm]
 \p_x V(+\infty,t)=0, \quad\quad 0<t<T,\\ [2mm]
  V(x,T)=\frac{1}{\gamma}(x+K)^{\gamma}, \quad\quad x>0.
\end{array} \right.
\ee

 We summarize the above results into the following theorem.

\begin{thm}\label{th1}\sl
In the following cases, problem \eqref{3.2} has trivial solution.
\begin{itemize}
\item[(1)] If \eqref{3.4} holds, the solution to problem \eqref{3.2} is $U(x+K)$.
\item[(2)] If \eqref{3.5} holds, the solution to problem \eqref{3.6} is the solution to problem \eqref{3.2} as well.
\end{itemize}
\end{thm}

 Recalling  \eqref{3.4} and \eqref{3.5}, in the following we always assume that
 \be\label{3.7}
 -r\gamma<\frac {a^2\gamma}{2(1-\gamma)}-r< 0.
 \ee
 In the case of \eqref{3.7}, there exists the free boundary.

\section{Dual transformation}
\setcounter{section}{3} \setcounter{equation}{0}

Define a dual transformation of $V(x,t)$ (see Pham (2009)) 
\begin{align}\label{4.1}
  v(y,t):=\max_{x>0}(V(x,t)-xy),\quad 0\leq y\leq y_0.
\end{align}
If $\partial_x V (\cdot,t)$ is strictly decreasing, which is equivalent to the strict concavity of $V (\cdot,t)$
(We will show this fact in Section 7), then the maximum in \eqref{4.1} will be attained at just one point
\begin{align}\label{4.2}
x=I(y,t),
\end{align}
which is the unique solution of
\begin{align}\label{4.3}
y=\partial_x V(x,t).
\end{align}
Using the coordinate transformation \eqref{4.2} yields
\begin{align}\label{4.4}
v(y,t)=[V(x,t)-x\partial_x V(x,t)]\Big|_{x=I(y,t)}=V(I(y,t),t)-yI(y,t).
\end{align}
Differentiating with respect to $y$ and $t$, we get
\be\label{4.5}
&&\hspace{-1cm}\partial_y v(y,t) =\partial_x V(I(y,t),t)\partial_y I(y,t)-y\partial_y I(y,t)-I(y,t)=-I(y,t), \\
&&\hspace{-1cm}\partial_{yy} v(y,t)=-\partial_y I(y,t)=-\frac 1{\partial_{xx} V(I(y,t),t)}, \label{4.6} \\
&&\hspace{-1cm}\p_t v(y,t)=\p_tV(I(y,t),t)+\partial_x V(I(y,t),t)\p_tI(y,t)-y\p_tI(y,t)=\p_tV(I(y,t),t). \label{4.7}
\ee
Substituting \eqref{4.5} into \eqref{4.4}, we have
\be\label{4.8}
V(I(y,t),t)=v(y,t)-y\partial_y v(y,t).
\ee
By the transformation \eqref{4.2} and \eqref{4.3}--\eqref{4.8}, the HJB equation in \eqref{3.2} becomes
\be\label{4.9}
\min\Big\{-\p_t v-\frac {a^2}{2}y^2\p_{yy}v+r v,\;\;v-y\p_y v- \frac 1\gamma (K-\p_y v)^\gamma\Big\}=0,\nonumber\\
\; 0<y<y_0,\;0<t<T.
\ee
Now we derive the terminal condition for $v(y,T)$. Note that
\be\label{4.10}
V(x,T)=\frac{1}{\gamma}(x+K)^{\gamma},
\ee
so $\partial_x V(x,T)=(x+K)^{\gamma-1},$ i.e., $[\partial_x V(x,T)]^{\frac 1{\gamma-1}}=x+K$. It follows that
\be\label{4.11}
y^{\frac 1{\gamma-1}}-K=x=I(y,T)=-\partial_y v(y,T),
\ee
and by \eqref{4.8}, we have
\be\label{4.12}
v(y,T)&=&V(I(y,T),T)+y\partial_y v(y,T)\nonumber\\
  &=&\frac{1}{\gamma}y^{\frac\gamma{\gamma-1}}
+y\Big(K-y^{\frac 1{\gamma-1}} \Big)\nonumber\\
  &=&\frac{1-\gamma}{\gamma}y^{\frac\gamma{\gamma-1}}+Ky .
\ee
Next, we determine the upper bound $y_0$ for $y$. In fact, $V(x,t)=\frac 1\gamma(x+K)^\gamma$ in the neighborhood of $x=0$,
so the upper bound is
 \be\label{4.13}
 y_0=\partial_x V(0,t)=K^{\gamma-1}.
 \ee
In addition, we need to determine the value $v(y_0,t)$. By \eqref{4.8}, we also have
 \be\label{4.14}
v(y_0,t)=V(0,t)+y_0\cdot 0=\frac 1\gamma K^\gamma.
 \ee

 Combining \eqref{4.9} and \eqref{4.12}--\eqref{4.14}, we obtain
 \be\label{4.15}
 \left\{\begin{array}{ll}
 \min\Big\{-\p_t v-\frac {a^2}{2}y^2\p_{yy}v+r v,\;\;v-y\p_y v- \frac 1\gamma (K-\p_y v)^\gamma\Big\}=0,\\
 \hspace{5cm} 0<y<K^{\gamma-1},\;\;0<t<T,\\
 v(K^{\gamma-1},t)=\frac 1\gamma K^\gamma,\;\;\;\;0<t<T,\\[2mm]
 v(y,T)=\frac{1-\gamma}{\gamma}y^{\frac\gamma{\gamma-1}}+Ky,\;\;\;\;0<y<K^{\gamma-1}.
 \end{array} \right.
 \ee
 In \eqref{4.15}, the equation is a linear parabolic equation,
 but the constraint condition
 \be\label{4.16}
 v-y\p_y v- \frac 1\gamma (K-\p_y v)^\gamma\geq 0
 \ee
 is very complicated. In the following section, we simplify this condition.

\noindent {\bf Remark:} The equation in \eqref{4.15} is degenerate on the boundary $y=0$.
According to Fichera's Theorem (see Oleinik-Radkevie (1973)), we must not put the boundary condition on $y=0$.

\section{Simplifying the complicated constraint condition}
\setcounter{section}{4} \setcounter{equation}{0}

Note that in the domain $\{(x,t)|V(x,t)=\frac 1\gamma(x+K)^\gamma\}$, we have
\be\label{5.1}
 \partial_x V(x,t)=(x+K)^{\gamma-1}, \quad \mbox{if } V(x,t)=\frac 1\gamma(x+K)^\gamma.
\ee
By the $y$ coordinate,
\be\label{5.2}
y=(K-\partial_y v)^{\gamma-1}, \quad \mbox{if } v-y\p_y v=\frac 1\gamma(K-\p_y v)^\gamma.
\ee
Deriving $\p_y v$ from the first equality in \eqref{5.2} yields
\be\label{5.3}
 \p_y v=K-y^{\frac 1{\gamma-1}},
\ee
and then substituting \eqref{5.3} into \eqref{4.16}, we have
\be\label{5.4}
 v\geq\frac{1-\gamma}{\gamma}y^{\frac\gamma{\gamma-1}}+Ky.
\ee
This is the simplified constraint condition. We assume that $u(y,t)$ satisfies
\be\label{5.5}
  \left\{\begin{array}{ll}
 \min\Big\{-\p_t u-\frac {a^2}{2}y^2\p_{yy}u+r u,\;\;u-\frac{1-\gamma}{\gamma}y^{\frac\gamma{\gamma-1}}-Ky\Big\}=0,\;\;\; (y,t)\in Q_y,\\[2mm]
 u(K^{\gamma-1},t)=\frac 1\gamma K^\gamma,\;\;\;\;0<t<T,\\[2mm]
 u(y,T)=\frac{1-\gamma}{\gamma}y^{\frac\gamma{\gamma-1}}+Ky,\;\;\;\;0<y<K^{\gamma-1},
  \end{array} \right.
 \ee
where
\bee
 Q_y=(0,\;K^{\gamma-1})\times(0,\;T).
 \eee
Moreover, we split the domain $Q_y$ into two parts, denote (see Fig. 1)
\be\label{5.6}
 && {\cal ER}_y=\Big\{u(y,t)=\frac{1-\gamma}{\gamma}y^{\frac\gamma{\gamma-1}}+Ky\Big\},\; \mbox{exercise region}, \\
 && {\cal CR}_y=\Big\{u(y,t)>\frac{1-\gamma}{\gamma}y^{\frac\gamma{\gamma-1}}+Ky\Big\},\; \mbox{continuation region}.
  \label{5.7}
\ee

\vspace{5cm}

\begin{center}
 \begin{picture}(650,-170)
 \thinlines \put(140,30){\vector(1,0){170}}
 \thinlines \put(140,120){\line(1,0){170}}
 \put(140,30){\vector(0,1){110}}
 \put(280,30){\line(0,1){90}}
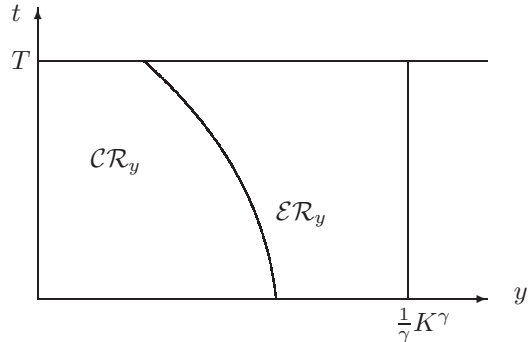
 \put(150,5){Fig. 1. ${\cal CR}_y$ and ${\cal ER}_y$}
 \put(130,135){$t$}\put(130,117){$T$}
 \qbezier(230,30)(225,80)(180,120)
 \put(320,30){$y$}
 \put(275,18){$\frac 1\gamma K^\gamma$}
 \put(230,60){${\cal ER}_y$}
 \put(160,80){${\cal CR}_y$}
 \end{picture}
 \end{center}

\begin{thm}\label{th2} \sl
The solution $u(x,t)$ to problem \eqref{5.5} is the solution to problem \eqref{4.15} as well.
\end{thm}

In order to prove this theorem, we first show the following two lemmas.

\begin{lem}\label{lem:1} \sl 
For any $(y,t) \in Q_y$, we have 
 \be\label{5.8}
 &&\p_y u= K-y^{\frac1{\gamma-1}},\;\;\;\;(y,t)\in {\cal ER}_y,\\
 &&\p_y u\leq K-y^{\frac1{\gamma-1}},\;\;\;\;(y,t)\in {\cal CR}_y.\label{5.9}
 \ee
\end{lem}

\noindent{\bf Proof:} Equation \eqref{5.8} follows from the definition \eqref{5.6} directly.
Also, in ${\cal CR}_y$
 \be\label{5.10}
 -\p_t u-\frac {a^2}{2}y^2\p_{yy}u+r u=0,\;\;\;\;(y,t)\in {\cal CR}_y.
 \ee
Differentiating \eqref{5.10} to $y$ yields
 \be\label{5.11}
 -\p_t(\p_y u)-\frac {a^2}{2}y^2\p_{yy}(\p_y u)-a^2y\p_{y}(\p_y u)+r (\p_y u)=0,\;\;\;\;(y,t)\in {\cal CR}_y.
 \ee
Note that
 \be\label{5.12}
 &&\p_y u(y,T)= K-y^{\frac1{\gamma-1}},\;\;\;\;0<y<K^{\gamma-1},\\
 &&\p_y u(y,t)= K-y^{\frac1{\gamma-1}},\;\;\;\;(y,t)\in \p({\cal CR}_y)\cap Q_y,\label{5.13}
 \ee
where $\p({\cal CR}_y)$ is the boundary of ${\cal CR}_y$.

Denote $w=k-y^{\frac 1{\gamma-1}}$, we further show that $w$ is a supersolution to problem \eqref{5.11}-\eqref{5.13} by
\bee
&&\p_y w=\frac 1{1-\gamma}y^{\frac 1{\gamma-1}-1}=\frac 1{1-\gamma}y^{\frac{2-\gamma}{\gamma-1}} \\
&&\p_{yy} w=-\frac{2-\gamma}{(1-\gamma)^2}y^{\frac 1{\gamma-1}-2}
\eee
and
\bee\label{5.12-1}
&&-\p_t w-\frac {a^2}{2}y^2\p_{yy} w-a^2y\p_{y}w+r w\\
&=\!\!\!&\frac {a^2}{2}\frac{2-\gamma}{(1-\gamma)^2}y^{\frac 1{\gamma-1}}-a^2\frac 1{1-\gamma}y^{\frac 1{\gamma-1}}
    +r(K-y^{\frac 1{\gamma-1}})\\
&=\!\!\!& rK+ \Big(\frac {a^2\gamma}{2(1-\gamma)^2}-r\Big) y^{\frac 1{\gamma-1}}>0,\;\;\;\;\;\;
\mbox{(by the first inequality in \eqref{3.7}).}
\eee
So $w$ is a supersolution of \eqref{5.11}-\eqref{5.13}. This means that \eqref{5.9} holds.
\hfill$\Box$

\begin{lem}\label{lem:2} \sl
The function
 \bee
 y\p_y u+ \frac 1\gamma (K-\p_y u)^\gamma
 \eee
 is increasing with respect to $\p_y u$ if $\p_y u\leq K-y^{\frac1{\gamma-1}}$.
\end{lem}

\noindent{\bf Proof:} Define a function
\bee
  f(z)=yz+ \frac 1\gamma (K-z)^\gamma,\;\;\;\;z\leq K-y^{\frac1{\gamma-1}}.
\eee
Then
\bee
f'(z)=y-( K-z)^{\gamma-1}\geq 0
\eee
if $z\leq K-y^{\frac1{\gamma-1}}$. 
\hfill$\Box$

\vspace{0.5cm}
\noindent{\bf Proof of Theorem \ref{th2}:}  Note that, from \eqref{5.5},
\be\label{5.14}
 &&-\p_t u-\frac {a^2}{2}y^2\p_{yy}u+r u\geq 0,\;\;\;\;(y,t)\in {\cal ER}_y,\\
 &&u=\frac{1-\gamma}{\gamma}y^{\frac\gamma{\gamma-1}}+Ky,\;\;\;\;\;\;(y,t)\in {\cal ER}_y.\label{5.15}
\ee
Rewrite \eqref{5.15} as
\be\label{5.16}
 u=y\Big(K-y^{\frac1{\gamma-1}}\Big)+\frac 1\gamma \Big(K-[K-y^{\frac1{\gamma-1}}]\Big)^\gamma,\;\;\;\;\;\;(y,t)\in {\cal ER}_y.
\ee
Applying \eqref{5.8} to \eqref{5.16}, we have
\be\label{5.17}
 u=y\p_y u+\frac 1\gamma (K-\p_y u)^\gamma,\;\;\;\;\;\;(y,t)\in {\cal ER}_y.
\ee
On the other hand, from \eqref{5.5}, in ${\cal CR}_y$
\be\label{5.18}
 &&-\p_t u-\frac {a^2}{2}y^2\p_{yy}u+r u= 0,\;\;\;\;(y,t)\in {\cal CR}_y,\\
 &&u\geq\frac{1-\gamma}{\gamma}y^{\frac\gamma{\gamma-1}}+Ky,\;\;\;\;\;\;(y,t)\in {\cal CR}_y.\label{5.19}
\ee
We rewrite \eqref{5.19} as
\be\label{5.20}
 u\geq y\Big(K-y^{\frac1{\gamma-1}}\Big)+\frac 1\gamma \Big(K-[K-y^{\frac1{\gamma-1}}]\Big)^\gamma,\;\;\;\;\;\;(y,t)\in {\cal CR}_y.
\ee
Applying \eqref{5.9} and Lemma \ref{lem:2}, we get
 \bee
 u\geq y\p_y u+\frac 1\gamma(K-\p_y u)^\gamma,\;\;\;\;\;\;(y,t)\in {\cal CR}_y.
 \eee
 \hfill$\Box$

\section{The free boundary of Problem \eqref{5.5}}
\setcounter{section}{5} \setcounter{equation}{0}

Denote
  \bee
 W^{2,1}_{p,loc}(Q_y)=\Big\{u(y,t):u,\;\p_y u,\;\p_{yy}u,\;\p_tu\in L^p(Q),
  \;\forall\; Q\subset\subset Q_y\Big\}.
   \eee
\begin{thm}\label{th3} \sl
The Problem \eqref{5.5} has a unique solution
$u\in W^{2,1}_{p,loc}(Q_y)\cap (\overline{Q_y}\;\backslash \{y=0\})$, and
 \be\label{7.1}
 &&\frac{1-\gamma}{\gamma}y^{\frac\gamma{\gamma-1}}+Ky\leq u(y,t)\leq e^{A(T-t)}\Big(\frac{1-\gamma}{\gamma}y^{\frac\gamma{\gamma-1}}+Ky\Big), \\
 &&\p_y \Big( u-\frac{1-\gamma}{\gamma}y^{\frac\gamma{\gamma-1}}-Ky\Big)\leq 0,\label{7.2}\\
 &&\p_t \Big( u-\frac{1-\gamma}{\gamma}y^{\frac\gamma{\gamma-1}}-Ky\Big)\leq 0,\label{7.3}
 \ee
 where $A=\frac{a^2}{2}\frac{\gamma}{(1-\gamma)^2}$.
\end{thm}

\noindent {\bf Proof:} According to the existence and uniqueness of $ W^{2,1}_{p,loc}(Q_y)\cap (\overline{Q_y}\;\backslash \{y=0\})$, the solution for system \eqref{5.5} can be proved by a standard penalty method (see Friedman (1975)). Here, we omit the details.
The first inequality in \eqref{7.1} follows from \eqref{5.5} directly, and now we prove the second inequality in \eqref{7.1}.
Denote
$$W(y,t):=e^{A(T-t)}\Big(\frac{1-\gamma}{\gamma}y^{\frac\gamma{\gamma-1}}+Ky\Big),$$
where $A>0$ to be determined later on. We first show that $W(y,t)$ is a supersolution to problem \eqref{5.5}. In fact,
\bee
&&-\p_t W-\frac {a^2}{2}y^2\p_{yy}W+rW\\
&&=Ae^{A(T-t)}\Big(\frac{1-\gamma}{\gamma}y^{\frac\gamma{\gamma-1}}+Ky\Big)
+e^{A(T-t)}\Big[\Big(-\frac{a^2}{2}\frac{1}{1-\gamma}+r\frac{1-\gamma}{\gamma}\Big)
  y^{\frac\gamma{\gamma-1}}+rKy\Big]\\
&&\geq e^{A(T-t)}\Big(A\frac{1-\gamma}{\gamma}-\frac{a^2}{2}\frac{1}{1-\gamma}\Big)y^{\frac\gamma{\gamma-1}}=0
\eee
if
\bee
A=\frac{a^2}{2}\frac{\gamma}{(1-\gamma)^2}.
\eee
So, $W(y,t)$ is a supersolution to problem \eqref{5.5}. Hence, the second inequality in \eqref{7.1} holds.

In addition, equation \eqref{7.2} follows from \eqref{5.8} and \eqref{5.9}. In order to prove  \eqref{7.3}, we define
$$w(y,t)=u(y,t-\delta)\;\;\;\; \mbox{for small } \delta>0.$$
From \eqref{5.5}, we know that $w(x,t)$ satisfies
\be\label{7.4}
  \left\{\begin{array}{ll}
 \min\Big\{-\p_t w-\frac {a^2}{2}y^2\p_{yy}w+r w,\;\;w-\frac{1-\gamma}{\gamma}y^{\frac\gamma{\gamma-1}}-Ky\Big\}=0,\;\;\; y>0,\;\delta<t<T,\\[2mm]
 w(K^{\gamma-1},t)=\frac 1\gamma K^\gamma,\;\;\;\;\delta<t<T,\\[2mm]
 w(y,T)=u(y,T-\delta)\geq \frac{1-\gamma}{\gamma}y^{\frac\gamma{\gamma-1}}+Ky,\;\;\;\;0<y<K^{\gamma-1}.
  \end{array} \right.
 \ee
Applying comparison principle to variational inequalities \eqref{5.5} and \eqref{7.4} with respect
 to terminal values (see Friedman (1982)), we obtain
 $$ u(y,t)\leq w(y,t)=u(y,t-\delta),\;\;\;\;y>0,\;\delta<t<T.$$
Thus $\p_t u\leq 0$ and \eqref{7.3} holds.
\hfill$\Box$

 Based on \eqref{7.2}, we define the free boundary
 $$h(t):=\min\Big\{y\Big| u(y,t)=\frac{1-\gamma}{\gamma}y^{\frac\gamma{\gamma-1}}+Ky\Big\},\;\;\;\;0\leq t<T.$$

\begin{thm}\label{th4} \sl
The free boundary function $h(t)$ is monotonic decreasing (Fig.2) with
\be\label{7.5}
 h(T):=\lim\limits_{t\rightarrow T^-}h(t)=\Big(\frac{rK}{\frac{a^2}{2}\frac{1}{1-\gamma}-r\frac{1-\gamma}{\gamma}}\Big)^{\gamma-1}.
\ee
Moreover, $h(t)\in C[0,T]\cap C^\infty[0,T)$.
\end{thm}

\noindent {\bf Proof:} First, from \eqref{7.3}, $h(t)$ is monotonic decreasing. 
Denote
$$
\varphi(y):=\frac{1-\gamma}{\gamma}y^{\frac\gamma{\gamma-1}}+Ky.
$$
In ${\cal ER}_y$, 
\bee
&&-\p_t \varphi-\frac {a^2}{2}y^2\p_{yy}\varphi+r\varphi
=\Big(-\frac{a^2}{2}\frac{1}{1-\gamma}+r\frac{1-\gamma}{\gamma}\Big)
  y^{\frac\gamma{\gamma-1}}+rKy\geq 0,
\eee
so
\bee
h(t)\geq\Big(\frac{rK}{\frac{a^2}{2}\frac{1}{1-\gamma}-r\frac{1-\gamma}{\gamma}}\Big)^{\gamma-1}, \quad 0\leq t<T.
\eee
Hence,
\bee
h(T)\geq\Big(\frac{rK}{\frac{a^2}{2}\frac{1}{1-\gamma}-r\frac{1-\gamma}{\gamma}}\Big)^{\gamma-1}.
\eee
In order to prove \eqref{7.5}, we suppose
\be
h(T)>\Big(\frac{rK}{\frac{a^2}{2}\frac{1}{1-\gamma}-r\frac{1-\gamma}{\gamma}}\Big)^{\gamma-1},
\ee
then it is not hard to get
\begin{align*}
\p_t u(y,T)>0, \quad \mbox{for } h(T)<y<\Big(\frac{rK}{\frac{a^2}{2}\frac{1}{1-\gamma}-r\frac{1-\gamma}{\gamma}}\Big)^{\gamma-1},
\end{align*}
which is a contradiction to \eqref{7.3}. Therefore, the desired result \eqref{7.5} holds.

Finally, the proof of $h(t)\in C[0,T]\cap C^\infty[0,T)$ is similar to the result in Friedman (1975). Here, we omit the details.
\hfill$\Box$

\vspace{4cm}

\begin{center}
 \begin{picture}(650,-170)
 \thinlines \put(140,30){\vector(1,0){170}}
 \thinlines \put(140,120){\line(1,0){170}}
 \put(140,30){\vector(0,1){110}}
 \put(280,30){\line(0,1){90}}
 \put(130,-5){Fig. 2. $y=h(t),\; \varphi(y)=\frac{1-\gamma}{\gamma}y^{\frac\gamma{\gamma-1}}+Ky$}
 \put(130,135){$t$}\put(130,117){$T$}
 \qbezier(230,30)(225,70)(180,120)
 \put(320,30){$y$}
 \put(275,18){$\frac 1\gamma K^\gamma$}
 \put(230,60){${\cal ER}_y$}
 \put(230,80){$u=\varphi(y)$}
 \put(160,70){${\cal CR}_y$}
 \put(160,50){$u>\varphi(y)$}
 \end{picture}
 \end{center}

\begin{thm}\label{th5} \sl
For any $(y,t)\in Q_y$, we have 
\be\label{7.7}
\p_{yy}u(y,t)>0.
\ee
\end{thm}

\noindent {\bf Proof:} If $(y,t)\in  {\cal ER}_y$, then $u=\frac{1-\gamma}{\gamma}y^{\frac\gamma{\gamma-1}}+Ky$. Thus, 
\begin{align*}
\p_{yy}u=\frac 1{1-\gamma}y^{\frac 1{\gamma-1}-1}>0, \quad (y,t)\in  {\cal ER}_y.
\end{align*}
If $(y,t)\in  {\cal CR}_y$, then
\be\label{7.8}
-\p_t u-\frac {a^2}{2}y^2\p_{yy}u+r u=0, \quad (y,t)\in  {\cal CR}_y.
\ee
Differentiating \eqref{7.8} with respect to $y$ twice yields
\be\label{7.9}
 -\p_t(\p_{yy} u)-\frac {a^2}{2}y^2\p_{yy}(\p_{yy} u)
 -a^2y\p_{y}(\p_{yy} u)+(r-a^2) (\p_{yy} u)=0,\;\;\;(y,t)\in  {\cal CR}_y.
\ee
Note that
\bee
 \p_{yy} u(y,t)>0, \quad t=T \mbox{ or } y=h(t).
\eee
Applying the minimum principle, we obtain
\begin{align*}
\p_{yy}u=\frac 1{1-\gamma}y^{\frac 1{\gamma-1}-1}>0,\;\;\;\;(y,t)\in  {\cal CR}_y.
\end{align*}
\hfill$\Box$

\noindent {\bf Remark:} From \eqref{4.6}, we have $\partial_{xx} V<0,$ which means $V$ is strict concave to $x$.

\section{The free boundary of original problem \eqref{3.2}}
\setcounter{section}{6} \setcounter{equation}{0}

Recalling on the free boundary $y=h(t)$
\be\label{8.1}
&&u(y,t)=\frac{1-\gamma}{\gamma}y^{\frac\gamma{\gamma-1}}+Ky, \quad y=h(t), \\
&&\p_yu(y,t)=-y^{\frac1{\gamma-1}}+K, \quad y=h(t).\label{8.2}
\ee
From dual transformation \eqref{4.2} and \eqref{4.5}, we know
\be\label{8.3}
x=-\p_yu(y,t).
\ee
Denote the free boundary of \eqref{3.2} by $x=g(t)$. Applying \eqref{8.2} and \eqref{8.3} yields 
\be\label{8.4}
&&g(t)=-\p_yu(h(t),t)=h(t)^{\frac1{\gamma-1}}-K.
\ee
Moreover,
\be
&&g'(t)=\frac1{\gamma-1}h(t)^{\frac1{\gamma-1}-1}h'(t)>0,\label{8.5}\\
&&g(T)=h(T)^{\frac1{\gamma-1}}-K= \frac{rK}{\frac{a^2}{2}\frac{1}{1-\gamma}-r\frac{1-\gamma}{\gamma}}-K,
\quad\quad (\mbox{by } \eqref{7.5}). \label{8.6}
\ee
Thus, we have following theorem.

\begin{thm}\label{th6} \sl
The free boundary $x=g(t)$ of problem \eqref{3.2} is monotonic increasing
(Fig 3) and $g(T)$ is determined by \eqref{8.6}.
 Moreover, $g(t)\in C[0,T]\cap C^\infty[0,T)$.
\end{thm}

\vspace{1.3cm}

\begin{center}
 \begin{picture}(650,90)
 \thinlines \put(140,30){\vector(1,0){170}}
 \thinlines \put(140,120){\line(1,0){160}}
 \put(140,30){\vector(0,1){110}}
 \put(150,5){Fig. 3. $x=g(t)$}
 \put(130,135){$t$}\put(130,117){$T$}
 \qbezier(210,30)(212,70)(240,120)
 \put(320,30){$x$}
 \put(255,85){${\cal CR}_x$}
 \put(230,60){$V>\frac 1\gamma (x+K)^\gamma$}
 \put(165,65){${\cal ER}_x$}
 \put(145,90){$V=\frac 1\gamma (x+K)^\gamma$}
 \end{picture}
 \end{center}

\noindent {\bf Financial meanings:} At time $t$, the manager should continue to invest according to
\eqref{eq:pi} if $x>g(t)$, while the investor should stop investment if $x<g(t)$.

\section{Concluding remark}
\setcounter{section}{7} \setcounter{equation}{0}

We explore a class of optimal investment problems mixed with optimal stopping in the financial investment.
The corresponding HJB equation, a free boundary problem of a fully nonlinear equation, is posed.
By means of a dual transformation, we obtain a new free boundary problem with a linear equation under a complicated
constraint condition. The key step is to simplify this complicated constraint condition.
In this way we study the properties of the free boundary and optimal strategy for investors.


\end{document}